\documentclass[preprintnumbers,amsmath,amssymb,nofootinbib]{revtex4}

\usepackage{graphicx}
\usepackage{epsfig}
\usepackage{color}

\newcommand{\beq}{\begin{equation}}
\newcommand{\eeq}{\end{equation}}
\newcommand{\bea}{\begin{eqnarray}}
\newcommand{\eea}{\end{eqnarray}}
\newcommand{\bwd}{\begin{widetext}}
\newcommand{\ewd}{\end{widetext}}

\begin{document}

\title{Three-dimensional envelope instability model in periodic focusing channels}


\author{Ji Qiang}
\email{jqiang@lbl.gov}
\affiliation{Lawrence Berkeley National Laboratory, Berkeley, CA 94720, USA}

\begin{abstract}
The space-charge driven envelope instability can be of great 
danger in high intensity accelerators and
	was studied using a two-dimensional (2D) envelope model and three-dimensional (3D) 
macroparticle simulations before.
	In this paper, we propose a three-dimensional envelope instability
model to study the instability for a bunched beam
	in a periodic solenoid and radio-frequency (RF) focusing channel and a periodic quadrupole and RF focusing channel.
This study shows that when the transverse zero current phase advance 
is below $90$ degrees, the beam envelope can still become unstable if
the longitudinal zero current phase advance is beyond $90$ degrees.
For the transverse zero current phase advance beyond $90$ degrees, 
the instability stopband width becomes larger with the increase of 
the longitudinal focusing strength and even shows different
structure from the 2D case 
when the longitudinal zero current phase advance is
beyond $90$ degrees. 
Breaking the symmetry of two longitudinal focusing RF cavities 
and the symmetry between the horizontal
focusing and the vertical focusing in the transverse plane in the
periodic quadrupole and RF channel makes the instability stopband broader.
This suggests that a
more symmetric accelerator lattice design might help reduce the 
range of the envelope instability in parameter space.

\end{abstract}

\maketitle

\section{Introduction}
The envelope instability as a space-charge driven collective 
instability presents a potentially great danger in high intensity
accelerators by causing beam size blow up
and quality degradation.
It has been studied theoretically~\cite{ingo1,jurgen,ingo3,davidson,okamoto,fedotov0,fedotov,lund} and 
experimentally~\cite{tiefenback,gilson,groening} since 1980s. 
In recent years, there was growing interest in further understanding
this instability and other structural resonances~\cite{jeon1,fukushima,li0,li,ingo2,jeon2,oliver,ingoprab17,ito,yuan,ingo4}.
Some of those studies were summarized in a recently published monograph~\cite{ingobook}.
However, most of those theoretical studies were based on a two-dimensional
model. Three-dimensional macroparticle simulations were carried out
for a bunched beam under the guidance of the two-dimensional
envelope instability model~\cite{ingo2,ingoprab17}.
It was found in reference~\cite{ingo2} that the instability stopband from the 
3D macroparticle simulation
is broader than that from the 2D envelope model.
Furthermore, the effect of the longitudinal synchrotron motion has not been systematically
studied in those macroparticle simulations and is missed in 
the 2D envelope instability model. 
In this paper, we proposed a three-dimensional
envelope instability model in periodic focusing channels.
Such a model can be used to systematically study the effect of longitudinal
synchrotron motion on the instability stopband for a bunched beam.
It can also be used to explore the stability in a fully
3D parameter space and to provide guidance for 3D macroparticle simulations.

The organization of this paper is as follows: after the introduction, we 
review the 2D envelope instability model in Section II; we present
the 3D envelope instability model in Section III; we present numerical
study of the envelope instability in a periodic transverse solenoid
and longitudinal RF focusing channel in Section IV; 
we present numerical study of 
the envelope instability in a periodic transverse quadrupole and longitudinal
RF focusing channel in Section V; and draw conclusions in Section VI.

\section{Two-dimensional envelope instability model}

For a two-dimensional coasting beam subject to external periodic focusing
forces and linear space-charge forces, the two-dimensional envelope equations 
are given as\cite{kv,jurgen}:
\begin{eqnarray}
	\frac{d^2 X}{d s^2} + k_x^2(s) X - \frac{K/2}{X+Y} - \frac{\epsilon_x^2}{X^3} & = & 0 \\
	\frac{d^2 Y}{d s^2} + k_y^2(s) Y - \frac{K/2}{X+Y} - \frac{\epsilon_y^2}{Y^3} & = & 0 
	\label{2denv}
\end{eqnarray}
where $X$ and $Y$ are horizontal and vertical rms beam sizes
respectively, $k_x^2$ and $k_y^2$ represent the external periodic focusing forces, 
$\epsilon_x$ and $\epsilon_y$ denote unnormalized rms emittances, and 
$K$ is the generalized perveance associated with the space-charge strength 
given by:
\begin{eqnarray}
	 K & = & \frac{q I}{2 \pi \epsilon_0 p_0 v_0^2 \gamma_0^2}
\end{eqnarray}
where $I$ is the current of the beam, $q$ is the charge of the particle,
$\epsilon_0$ is the vacuum permittivity, $p_0$ is the momentum
of the reference particle, $v_0$ is the speed of the reference particle,
and $\gamma_0$ is the relativistic factor of the reference particle.

The above equations can be linearized with respect to a periodic solution
(i.e. matched solution) as:
\begin{eqnarray}
	X(s) & = & X_0(s) + x(s)  \\
	Y(s) & = & Y_0(s) + y(s) 
\end{eqnarray}
where $X_0$ and $Y_0$ denote the periodic matched envelope solutions 
and $x$ and $y$ denote small perturbations
\begin{equation}
	x(s) \ll X_0(s), \ \ \ \ y(s) \ll Y_0(s)
\end{equation}
The equations of motion for the small perturbations are given by:
\begin{eqnarray}
	\frac{d^2 x(s)}{d s^2} + a_1(s) x(s) + a_{12}(s) y(s) = 0  \\
	\frac{d^2 y(s)}{d s^2} + a_2(s) y(s) + a_{12}(s) x(s) = 0  
\end{eqnarray}
where
\begin{eqnarray}
	a_{12}(s) & = & 2K/(X_0(s) + Y_0(s))^2  \\
	a_1(s) & = & k_x^2(s) + 3 \epsilon_x^2/X_0^4(s) + a_{12}(s)  \\
	a_2(s) & = & k_y^2(s) + 3 \epsilon_y^2/Y_0^4(s) + a_{12}(s)  
\end{eqnarray}
With $\xi = (x,x',y,y')^T$, the prime denotes derivative with respect to $s$, 
and $T$ denotes the transpose of a matrix, 
the above equations can be rewritten in matrix notation as:
\begin{eqnarray}
	\frac{d \xi}{d s} & = & A_4(s) \xi(s)
	\label{2deq}
\end{eqnarray}
with the periodic matrix
\begin{eqnarray}
	A_4(s) & = & \begin{pmatrix}
		0 & 1 & 0 & 0 \\
		-a_1(s) & 0 & -a_{12}(s) & 0 \\
		0 & 0 & 0 & 1 \\
		-a_{12}(s) & 0 & -a_2(s) & 0 \\
		 \end{pmatrix}
\end{eqnarray}
Let $\xi(s) = M_4(s) \xi(0)$ be the solution of above equation,
substituting this equation into 
Eq.~\ref{2deq} results in
\begin{eqnarray}
	\frac{d M_4(s)}{ds} & = & A_4(s) M_4(s)  
\end{eqnarray}
where $M_4(s)$ denotes the $4 \times 4$ transfer matrix solution of $\xi(s)$ and
$M_4(0)$ is a $4\times 4$ unit matrix. The matrix $A_4(s)$ is a periodic function of $s$
with a length of period $L$. Following the Floquet's theorem, the
solution of $M_4(s)$ after $n$ lattice periods can be written as
\begin{eqnarray}
	M_4(s+nL) & = & M_4(s)M_4(L)^n 
\end{eqnarray}
This matrix solution will remain finite as $n->\infty$, only if all amplitudes 
of the eigenvalues of the matrix $M_4(L)$ be less than or equal to one.
Since the matrix $M_4(L)$ is a real symplectic matrix, the eigenvalues
of the matrix occur both as reciprocal and as complex-conjugate pairs.
Therefore, for stable solutions, 
all eigenvalues of the matrix $M_4(L)$ have to lie on a
unit circle in the complex plane. 
The eigenvalues of the matrix $M_4(L)$ can be expressed in polar coordinates as:
\begin{eqnarray}
	\lambda & = & |\lambda| \exp{(i \phi)}
\end{eqnarray}
where the amplitude $|\lambda|$ of the eigenvalue gives the growth rate
(or damping rate) of the envelope eigenmode through one lattice period and the
phase shift $\phi$ of the eigenvalue gives the phase of the envelope mode
oscillation through one period.
For an unstable envelope mode, there are two possibilities~\cite{jurgen}:
\begin{enumerate}
	\item one or both eigenvalue pairs lie on the real axis: $\phi_{1,2}=180^\circ$,
	\item the phase shift angles are equal: $\phi_1 = \phi_2$.
\end{enumerate}
The first case can be seen as a half-integer parametric resonance between
the focusing lattice and the envelope oscillation mode. The second
case is a confluent resonance between two envelope oscillation modes since they
have the same oscillation frequencies. 

\section{Three-dimensional envelope instability model}
The 3D envelope equations have been used to study the halo particle formation
mechanism (e.g. particle-core model) for a bunched beam in high intensity 
accelerators~\cite{bongardt,allen,qiang0,comunian}.
There, the mismatched envelope oscillation resonates with a test particle and
drives the particle into large amplitude becoming a halo particle.
The mismatched envelope oscillation itself is stable in that case. 
In this paper, 
we study the stability/instability of the mismatched envelope 
oscillation itself in periodic focusing channels.

For a 3D uniform density ellipsoidal beam inside a periodic focusing channel without acceleration, 
the three-dimensional envelope equations are given as~\cite{sacherer,ryne}:
\begin{eqnarray}
	\frac{d^2 X}{d s^2} + k_x^2(s) X - I_x(X,Y,Z)X - \frac{\epsilon_x^2}{X^3} & = & 0 \\
	\frac{d^2 Y}{d s^2} + k_y^2(s) Y - I_y(X,Y,Z)Y - \frac{\epsilon_y^2}{Y^3} & = & 0   \\
	\frac{d^2 Z}{d s^2} + k_z^2(s) Z - I_z(X,Y,Z)Z - \frac{(\epsilon_z/\gamma^2)^2}{Z^3} & = & 0 
	\label{3denv}
\end{eqnarray}
with
\begin{eqnarray}
	I_i(X,Y,Z) = C\int_0^{\infty} \frac{dt}{(e_i^2+t)\sqrt{(X^2+t)(Y^2+t)(\gamma^2 Z^2+t)}}
\end{eqnarray}
where $X$, $Y$, and $Z$ are horizontal, vertical, and longitudinal rms beam 
sizes respectively, $e_i = X, Y, \gamma Z$, for $i=x,y,z$, and $C = \frac{1}{2}\frac{3}{4\pi \epsilon_0}\frac{q}{mc^2}\frac{I}{f_{rf} \beta^2 \gamma^2}\frac{1}{5\sqrt{5}}$. Here, $\epsilon_0$ is the
vacuum permittivity, $q$ the charge, $mc^2$ the rest energy of the particle, $c$ the light speed in vacuum, $I$ the average beam current, $f_{rf}$ the RF
bunch frequency, $\beta = v/c$, $v$ the bunch velocity, and the relativistic
factor $\gamma = 1/\sqrt{1-\beta^2}$.

The above equations can be linearized with respect to periodic solutions
(i.e. matched solutions) as:
\begin{eqnarray}
	X(s) & = & X_0(s) + x(s)  \\
	Y(s) & = & Y_0(s) + y(s)  \\
	Z(s) & = & Z_0(s) + z(s)  
\end{eqnarray}
where $X_0$, $Y_0$ and $Z_0$ denote the periodic matched envelope solutions 
and $x$, $y$ and $z$ denote small perturbations
\begin{equation}
	x(s) \ll X_0(s), \ \ \ \ y(s) \ll Y_0(s), \ \ \ \ z(s) \ll Z_0(s)
\end{equation}
The equations of motion for these small perturbations are given by:
\begin{eqnarray}
	\frac{d^2 x}{d s^2} + a_1(s) x(s) + a_{12}(s) y(s) + \gamma^2 a_{13}(s) z(s) = 0  \\
	\frac{d^2 y}{d s^2} + a_{12}(s) x(s) + a_2(s) y(s)  +\gamma^2 a_{23}(s) z(s) = 0   \\
	\frac{d^2 z}{d s^2} + a_{13}(s) x(s) + a_{23}(s) y(s)  + a_3(s) z(s) = 0  
\end{eqnarray}
where
\begin{eqnarray}
	a_1(s) & = &  k_x^2 + 3 \epsilon_x^2/X_0^4 - I_{x}(X_0,Y_0,Z_0) + 3 X_0^2 F_{xx} \\
	a_{12}(s) & = & X_0 Y_0 F_{xy} \\
	a_{13}(s) & = & X_0 Z_0 F_{xz} \\
	a_2(s) & = &  k_y^2 + 3 \epsilon_y^2/Y_0^4 - I_{y}(X_0,Y_0,Z_0) + 3 Y_0^2 F_{yy} \\
	a_{23}(s) & = & Y_0 Z_0 F_{yz} \\
	a_3(s) & = &  k_z^2 + 3 (\epsilon_z/\gamma^2)^2/Z_0^4 - I_{z}(X_0,Y_0,Z_0) + 3 \gamma^2 Z_0^2 F_{zz} 
\end{eqnarray}
where 
\begin{eqnarray}
	F_{xx} & = & C\int_0^{\infty} (X_0^2+t)^{-5/2}(Y_0^2+t)^{-1/2}(Z_0^2 \gamma^2+t)^{-1/2} dt  \\
	F_{xy} & = & C\int_0^{\infty} (X_0^2+t)^{-3/2}(Y_0^2+t)^{-3/2}(Z_0^2 \gamma^2+t)^{-1/2} dt  \\
	F_{xz} & = & C\int_0^{\infty} (X_0^2+t)^{-3/2}(Y_0^2+t)^{-1/2}(Z_0^2 \gamma^2+t)^{-3/2} dt  \\
	F_{yy} & = & C\int_0^{\infty} (X_0^2+t)^{-1/2}(Y_0^2+t)^{-5/2}(Z_0^2 \gamma^2+t)^{-1/2} dt  \\
	F_{yz} & = & C\int_0^{\infty} (X_0^2+t)^{-1/2}(Y_0^2+t)^{-3/2}(Z_0^2 \gamma^2+t)^{-3/2} dt  \\
	F_{zz} & = & C\int_0^{\infty} (X_0^2+t)^{-1/2}(Y_0^2+t)^{-1/2}(Z_0^2 \gamma^2+t)^{-5/2} dt  
\end{eqnarray}

With $\xi = (x,x',y,y',z,z')^T$, 
the above equations can be rewritten in matrix notation as:
\begin{eqnarray}
	\frac{d \xi}{d s} & = & A_{6}(s) \xi(s)
	\label{3deq}
\end{eqnarray}
with the periodic matrix
\begin{eqnarray}
	A_{6}(s) & = & \begin{pmatrix}
		0 & 1 & 0 & 0 & 0 & 0 \\
		-a_1(s) & 0 & -a_{12}(s) & 0 & -\gamma^2 a_{13}(s) & 0 \\
		0 & 0 & 0 & 1 & 0 & 0 \\
		-a_{12}(s) & 0 & -a_2(s) & 0 & -\gamma^2 a_{23}(s) & 0 \\
		0 & 0 & 0 & 0 & 0 & 1 \\
		-a_{13}(s) & 0 & -a_{23}(s) & 0 & -a_{3}(s) & 0 \\
		 \end{pmatrix}
\end{eqnarray}
Let $\xi(s) = M_{6}(s) \xi(0)$, substituting this equation into Eq.~\ref{3deq}
results in
\begin{eqnarray}
	\frac{d M_6(s)}{ds} & = & A_6(s) M_6(s)  
\end{eqnarray}
where $M_6(s)$ denotes the $6\times 6$ transfer matrix solution of 
$\xi(s)$ and $M_6(0)$ is a $6\times 6$ unit matrix. 
The above ordinary differential equation can be solved using the matched
envelope solutions and numerical integration.
Similar to the 2D envelope instability model, the stability of these envelope 
perturbations is determined by the eigenvalues of the transfer matrix $M_6(L)$
through one lattice period.
For the envelope oscillation to be stable, 
all eigenvalues of the $M_6(L)$ have to stay on the unit circle.
The amplitude of the eigenvalue gives the envelope mode growth 
(or damping) rate through one lattice period, while the phase of 
the eigenvalue yields the 
mode oscillation frequency. When the amplitude of any eigenvalue is
greater than one, the envelope oscillation becomes unstable.


\section{Envelope Instability in a periodic solenoid and RF channel}

\begin{figure}[!htb]
   \centering
   \includegraphics*[angle=0,width=210pt]{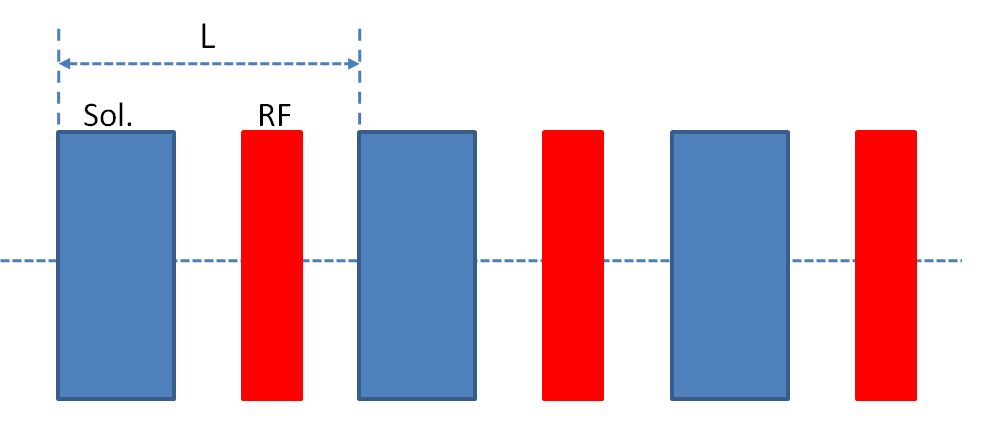}
	\caption{Schematic plot of a periodic solenoid and RF channel. }
   \label{sol3d}
\end{figure}
\begin{figure*}[!htb]
   \centering
   \includegraphics*[angle=0,width=210pt]{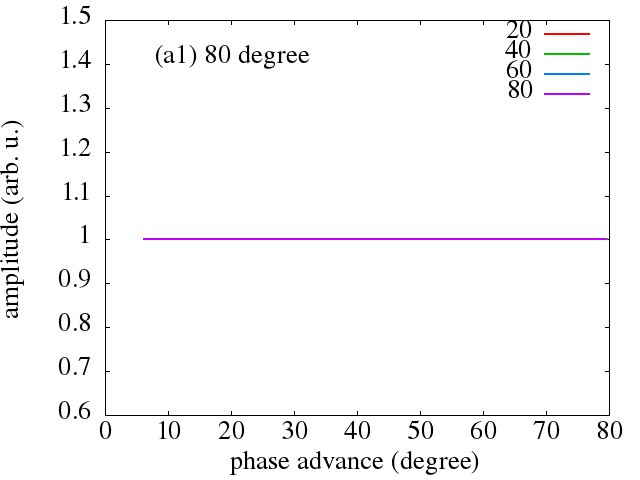}  
   \includegraphics*[angle=0,width=210pt]{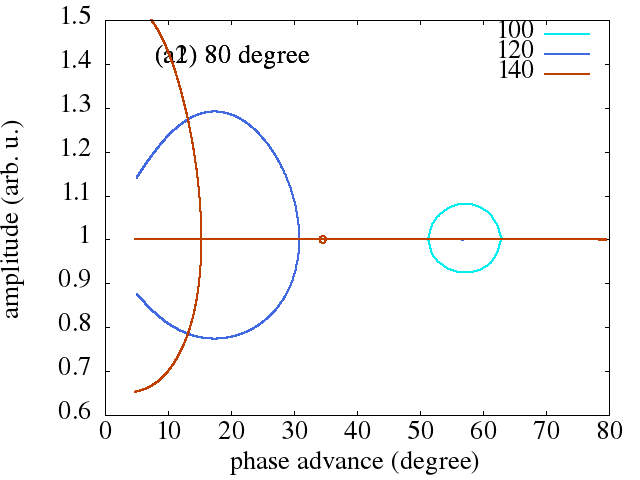}  
   \includegraphics*[angle=0,width=210pt]{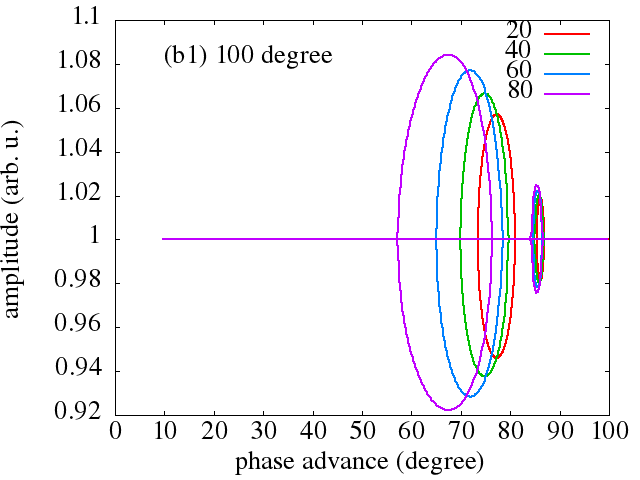}
   \includegraphics*[angle=0,width=210pt]{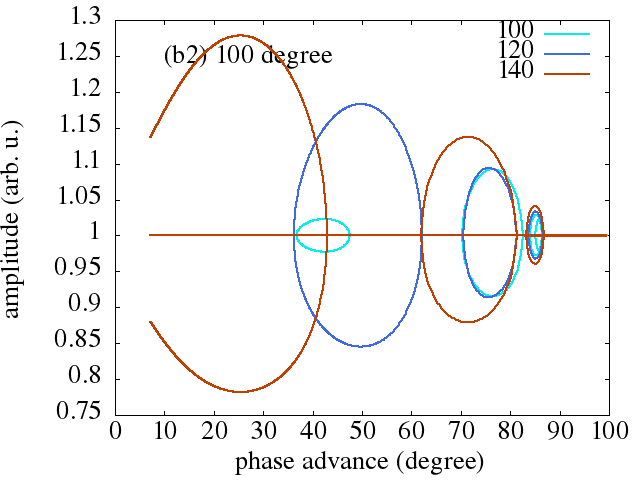}
   \includegraphics*[angle=0,width=210pt]{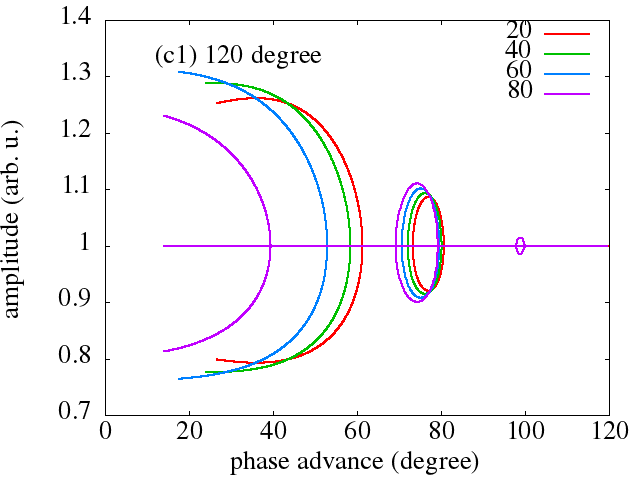}
   \includegraphics*[angle=0,width=210pt]{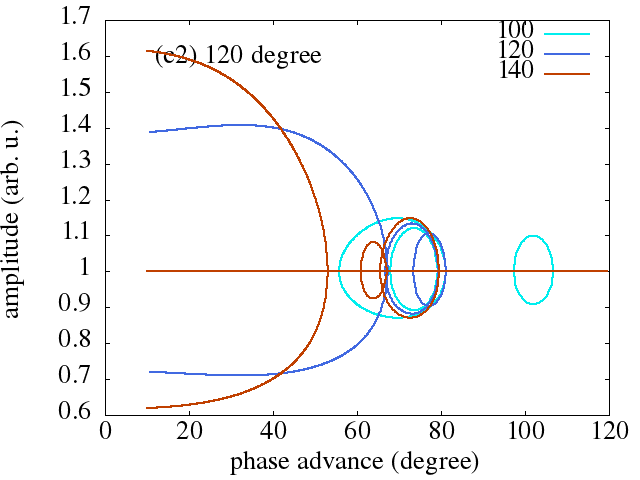}
   \includegraphics*[angle=0,width=210pt]{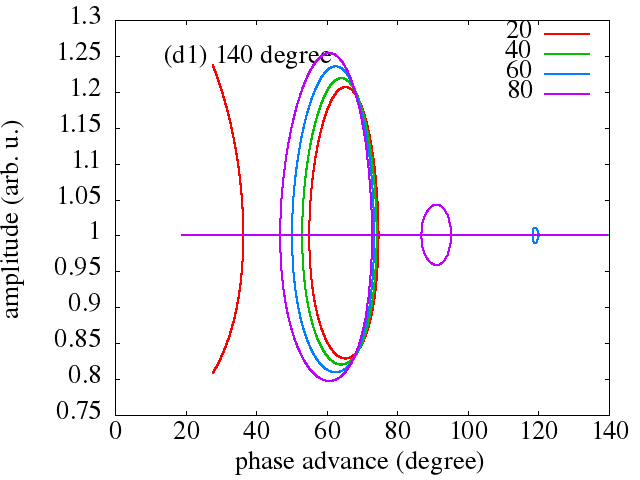}
   \includegraphics*[angle=0,width=210pt]{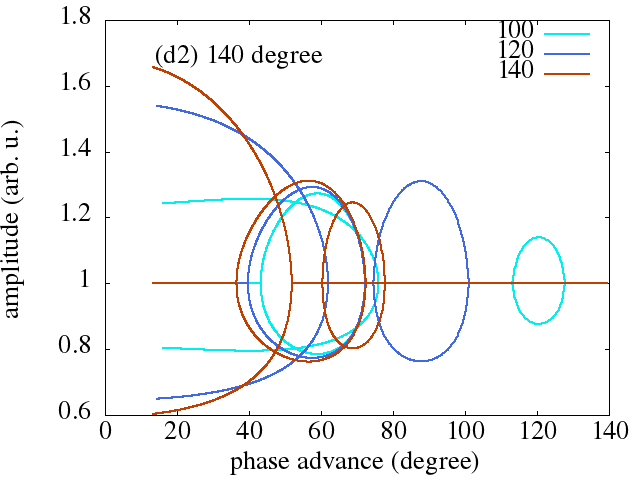}
   \caption{The 3D envelope mode amplitudes as a function of 
	depressed transverse phase advance with $20$, $40$,
	$60$, $80$, $100$, $120$, and $140$ degree zero current 
	longitudinal phase advances
	for (a) $80$ degree, (b) $100$ degree, (c) $120$ degree,
	and (d) $140$ degree zero current transverse phase advances in
	a periodic solenoid-RF channel.
	}
   \label{sol3damp}
\end{figure*}
\begin{figure*}[!htb]
   \centering
   \includegraphics*[angle=0,width=210pt]{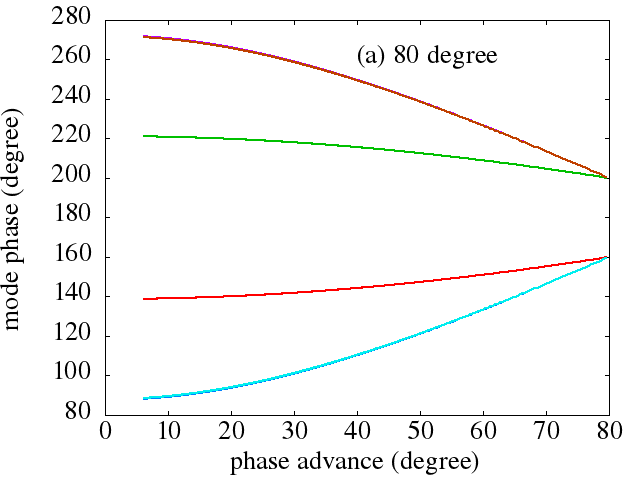} 
   \includegraphics*[angle=0,width=210pt]{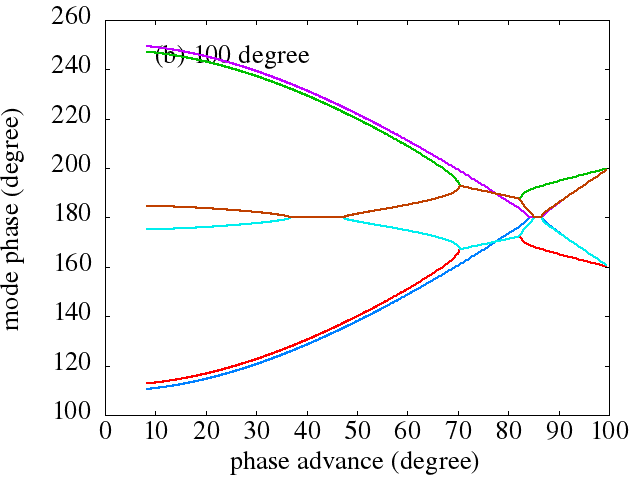} 
   \includegraphics*[angle=0,width=210pt]{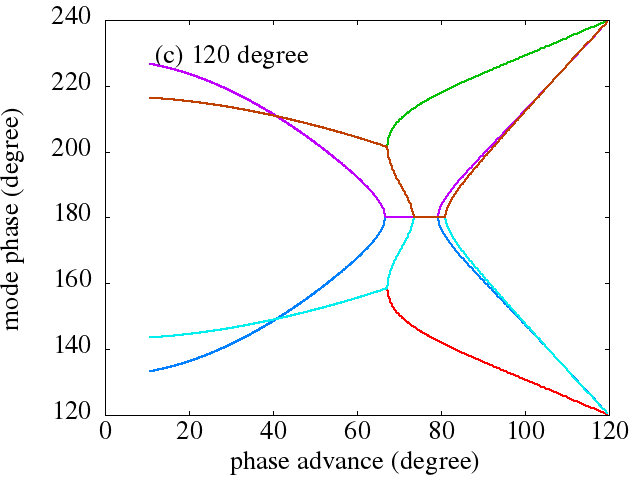} 
   \includegraphics*[angle=0,width=210pt]{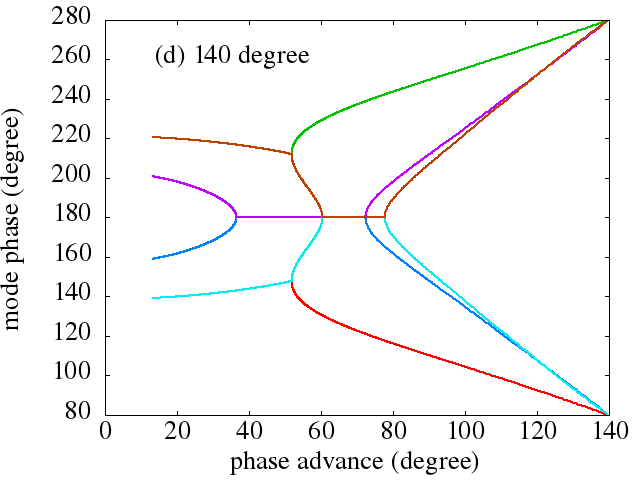} 
   \caption{The 3D envelope mode phases as a function of 
	depressed transverse phase advance with (a) 
	$80$ degree, (b) $100$ degree, (c) $120$ degree,
	and (d) $140$ degree zero current longitudinal
	and transverse phase advances in a periodic solenoid-RF channel.
	}
   \label{sol3dphase}
\end{figure*}
\begin{figure*}[!htb]
   \centering
   \includegraphics*[angle=0,width=210pt]{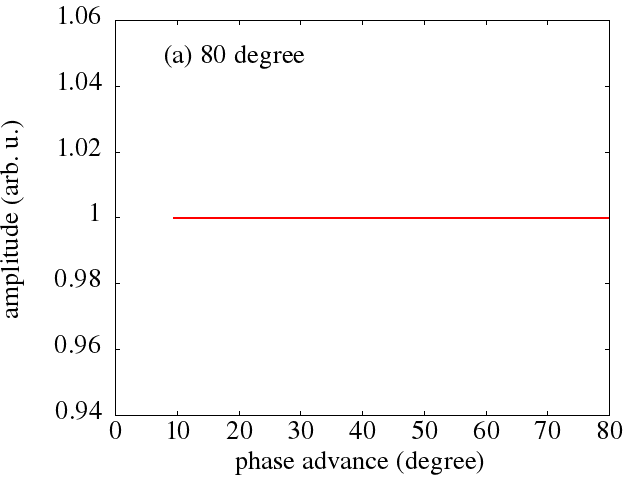}
   \includegraphics*[angle=0,width=210pt]{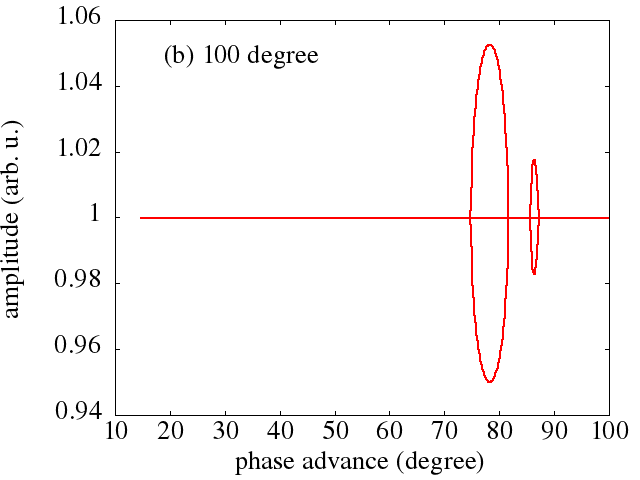}
   \includegraphics*[angle=0,width=210pt]{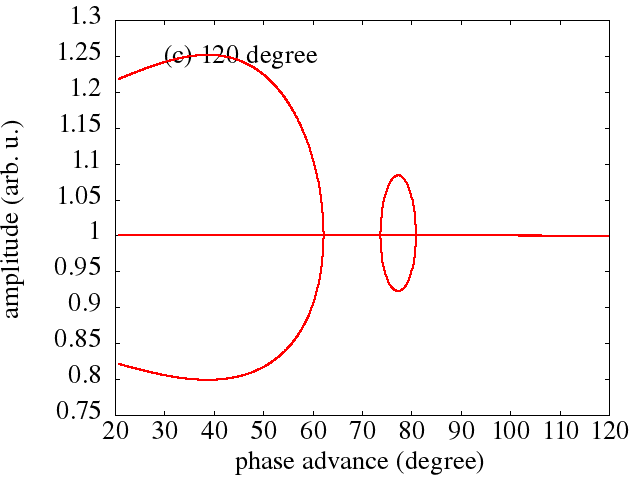}
   \includegraphics*[angle=0,width=210pt]{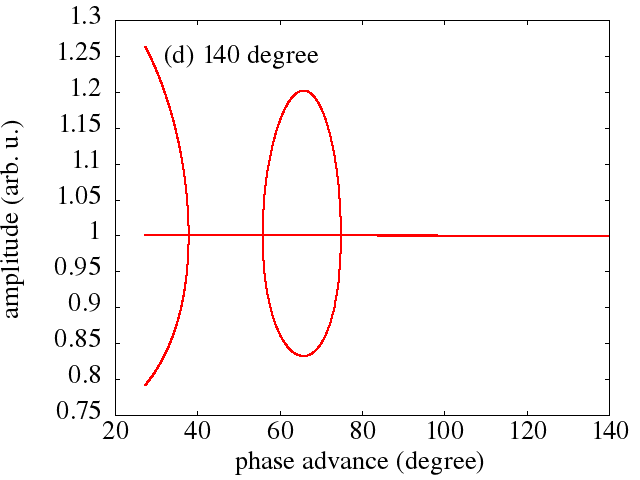}
   \caption{The 2D envelope mode amplitudes as a function of 
	depressed transverse phase advance
	for (a) $80$ degree, (b) $100$ degree, (c) $120$ degree,
	and (d) $140$ degree zero current transverse phase advances 
	in a periodic solenoid channel.  }
   \label{sol2damp}
\end{figure*}
\begin{figure*}[!htb]
   \centering
   \includegraphics*[angle=0,width=210pt]{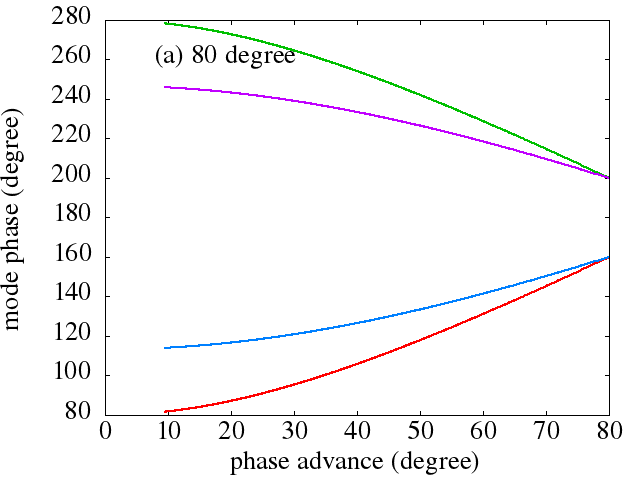}
   \includegraphics*[angle=0,width=210pt]{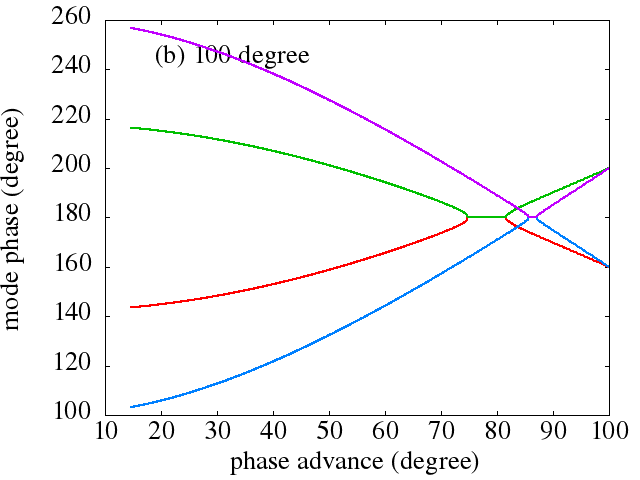}
   \includegraphics*[angle=0,width=210pt]{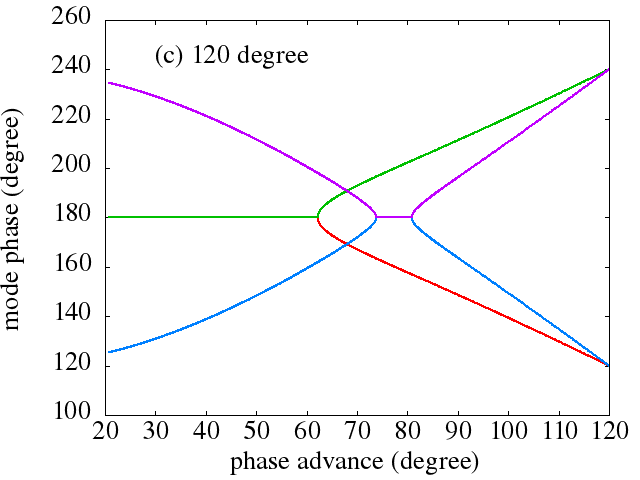}
   \includegraphics*[angle=0,width=210pt]{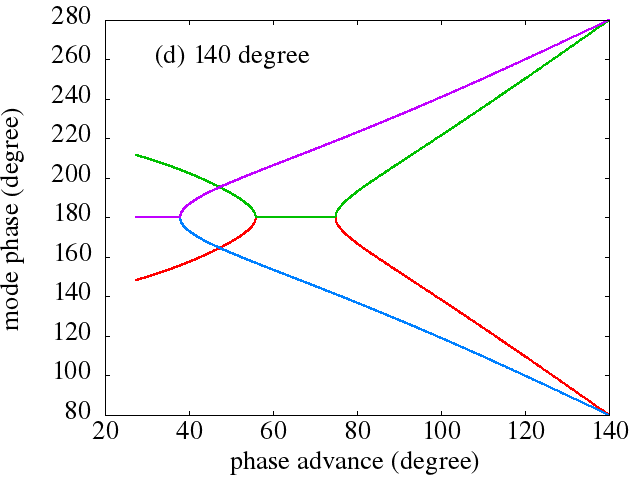}
   \caption{The 2D envelope mode phases as a function of 
	depressed transverse phase advance
	for (a) $80$ degree, (b) $100$ degree, (c) $120$ degree,
	and (d) $140$ degree zero current transverse phase advances 
	in a periodic solenoid channel.  }
   \label{sol2dphase}
\end{figure*}
\begin{figure}[!htb]
   \centering
   \includegraphics*[angle=0,width=210pt]{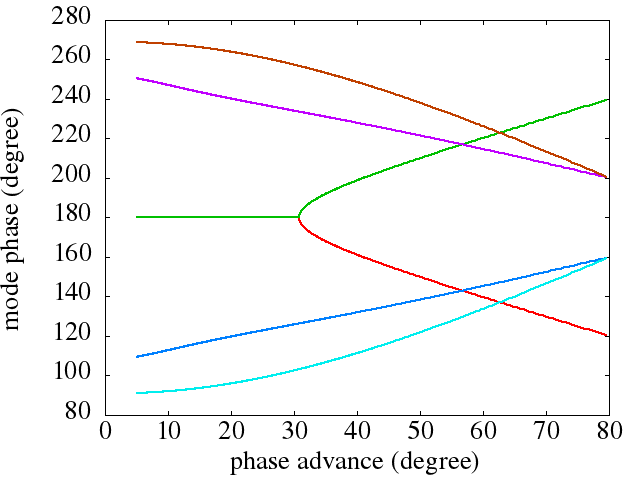} 
   \includegraphics*[angle=0,width=210pt]{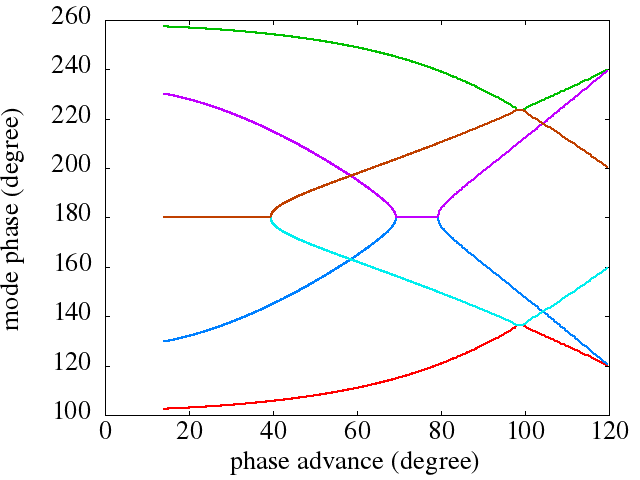} 
	  \caption{The 3D envelope mode phases as a function of 
the depressed transverse phase advance for zero current (a) 
	transverse $80$ degree and longitudinal $120$ degree, 
	(b) transverse $120$ degree and longitudinal $80$ degree
	phase advance in a periodic solenoid channel. }
   \label{sol80120phase}
\end{figure}

We first studied the envelope instability in a transverse solenoid focusing
and longitudinal RF focusing periodic channel.
A schematic plot of this periodic channel is shown in Fig.~\ref{sol3d}.
Each period of the channel consists of a $0.2$ meter solenoid and a $0.1$
meter RF bunching cavity.
The total length of the period is $0.5$ meters.
The proton bunch has a kinetic energy of $150$ MeV and normalized rms emittances
of $0.2$ um, $0.2$ um, and $0.2$ um in horizontal, vertical, and longitudinal
directions respectively.
Figures~\ref{sol3damp}-\ref{sol3dphase} show the 3D envelope mode amplitudes and phases
as a function of transverse depressed phase advance for different
zero current transverse and longitudinal phase advances.
As a comparison, we also show in Figs.~\ref{sol2damp}-\ref{sol2dphase}
the 2D envelope mode amplitudes and phases as a function of depressed 
transverse phase advance for the same zero current transverse phase advances. 
Here, the 2D periodic solenoid channel has the same length of period 
as the 3D channel. It is seen that in the 2D periodic solenoid channel, the envelope instability 
occurs when the zero current phase advance is over $90$ degrees. In the
3D periodic solenoid-RF channel, the envelope instability occurs even with
the zero current transverse phase advance $80$ degrees but longitudinal 
phase advance beyond $90$ degrees as shown in Fig.~\ref{sol3damp} (a2). 
There is no 
instability if both the transverse zero current phase advance and the longitudinal
zero current phase advance are below $90$ degrees as seen 
in Fig.~\ref{sol3damp} (a1).
For the 3D envelope modes, when the longitudinal zero current
phase advance below $90$ degrees and the transverse zero
current phase above $90$ degrees as shown in Figs~\ref{sol3damp} (b1, c1, and d1), 
the instability stopband
becomes broader as the zero current longitudinal phase advance increases.
This is probably because the longitudinal synchrotron motion
helps bring particles with different depressed transverse tunes into the resonance.
A faster synchrotron motion might result in more particles falling 
into the resonance 
and hence a broader instability stopband. 
For small longitudinal zero current phase advance (e.g. $20$
degrees), the 3D envelope mode show the stopband
similar to that of the 2D envelope mode.
When the longitudinal
zero current phase advance is above $90$ degrees, as shown in 
Fig.~\ref{sol3damp} (b2, c2, and d2), the 3D
envelope instability shows more complicated structure and larger instability stopband width
than the 2D envelope instability. 

In the 2D periodic transverse solenoid focusing channel, for a coasting beam
with equal horizontal and vertical emittances,
it is seen in Fig.~\ref{sol2dphase}, the envelope instabilities are due to
the $180$ degree half-integer parametric resonance. 
However, for a bunched beam, as shown in Fig.~\ref{sol3dphase}, 
besides the $180$ degree half-integer
resonance, there are also confluent resonances where two
envelope modes have the same frequencies and resonate with each other.
The existence of both instability mechanisms results in more complicated
structure as shown in Figs.~\ref{sol3damp} (b2, c2, and d2).

The 3D envelope instability shows asymmetry between the
transverse direction and the longitudinal direction in 
the 3D periodic solenoid
and RF channel. Figure~\ref{sol80120phase} shows the envelope mode phases as a 
function of depressed transverse phase advance for a case with zero current
$80$ degree transverse phase advance and $120$ degree longitudinal
phase advance, and a case with zero current $120$ degree transverse phase advance
and $80$ degree longitudinal phase advance. The envelope mode amplitudes 
for both cases are shown in Fig.~\ref{sol3damp} (a2 and c1).
For the $80$ degree zero current transverse
phase advance, there is
only one major unstable stopband below $30$ degree depressed transverse 
phase advance
due to half-integer parametric resonance as shown in the left plot
of Fig.~\ref{sol80120phase}.
For the $120$ degree zero current transverse phase advance, there are three 
unstable regions, two due to the half-integer parameter resonance and
one due to the confluent resonance as shown in the right plot of 
Fig.~\ref{sol80120phase}.
This asymmetry is probably related to the two degrees of fredom in the
transverse plane while only one in the longitudinal direction.

\section{Envelope Instability in a periodic quadrupole-RF channel}

Next, we studied the 3D envelope instability
in a periodic transverse quadrupole focusing and longitudinal RF focusing
channel for the same bunched proton beam. 
\begin{figure}[!htb]
   \centering
   \includegraphics*[angle=0,width=300pt]{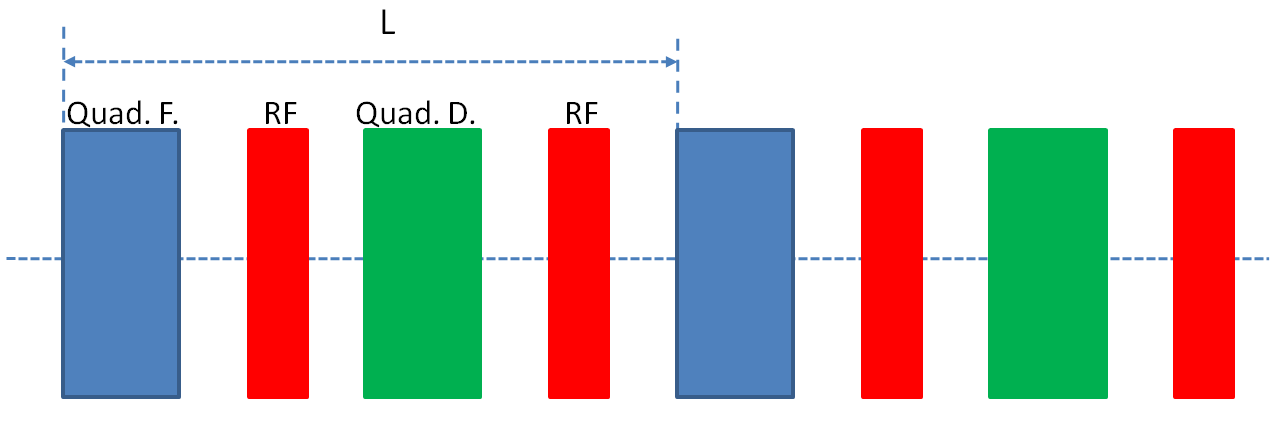}
	\caption{Schematic plot of a periodic quadrupole and RF channel. }
   \label{fd3d}
\end{figure}
\begin{figure*}[!htb]
   \centering
   \includegraphics*[angle=0,width=210pt]{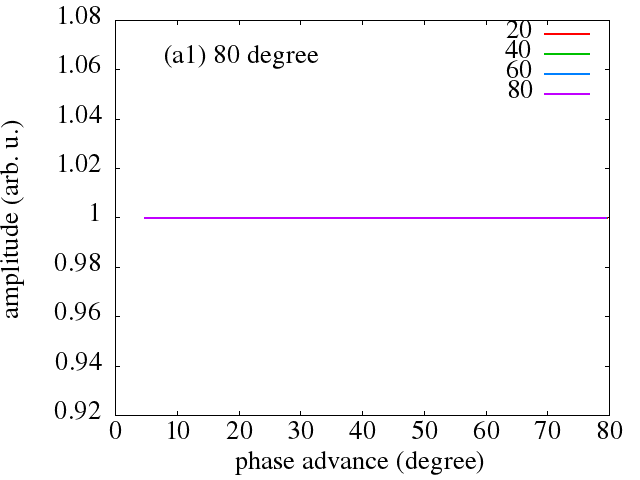} 
   \includegraphics*[angle=0,width=210pt]{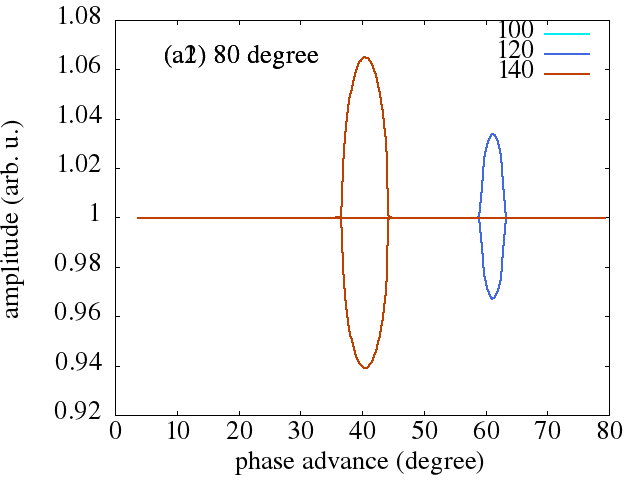} 
   \includegraphics*[angle=0,width=210pt]{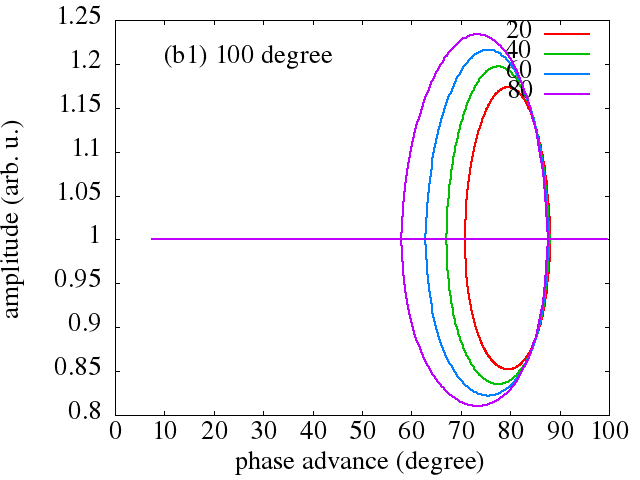}
   \includegraphics*[angle=0,width=210pt]{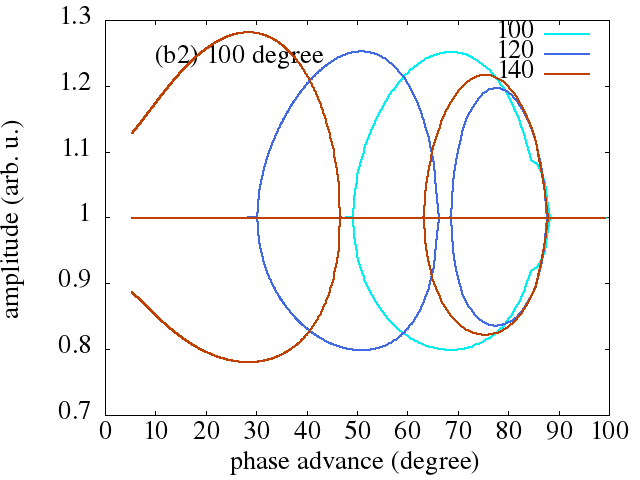}
   \includegraphics*[angle=0,width=210pt]{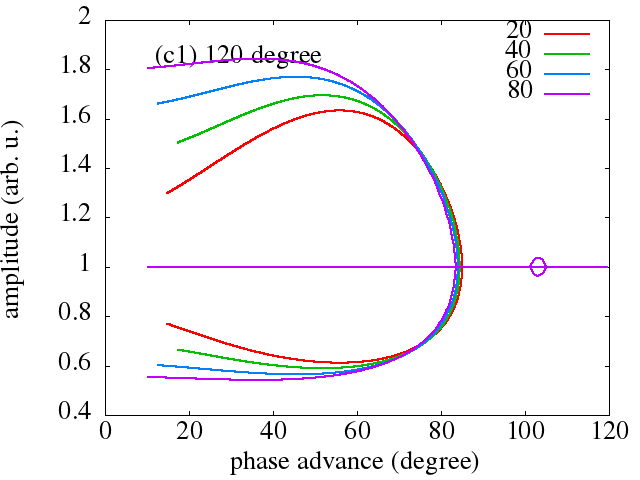}
   \includegraphics*[angle=0,width=210pt]{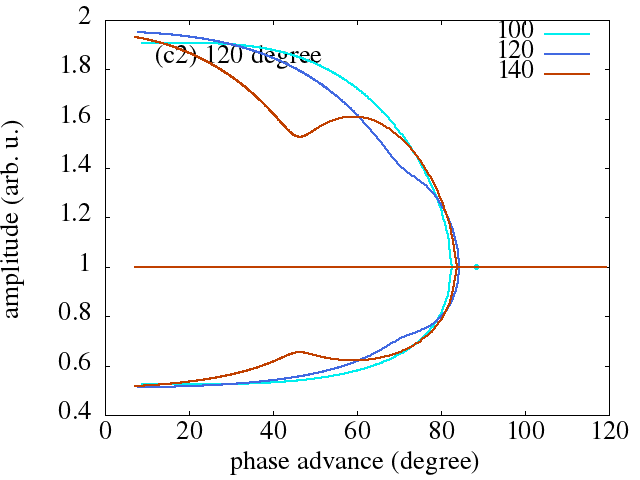}
   \includegraphics*[angle=0,width=210pt]{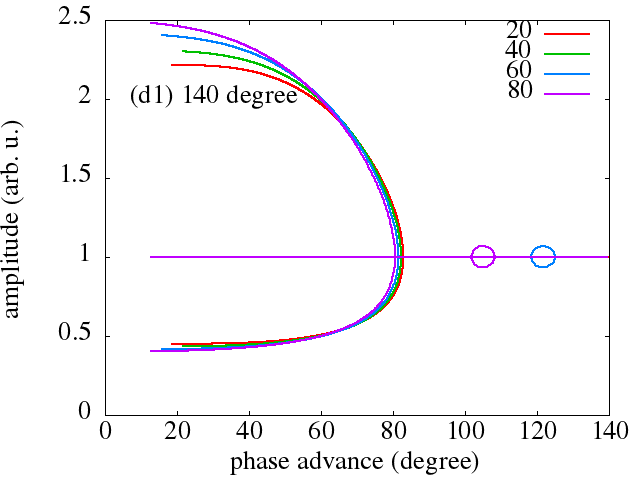}
   \includegraphics*[angle=0,width=210pt]{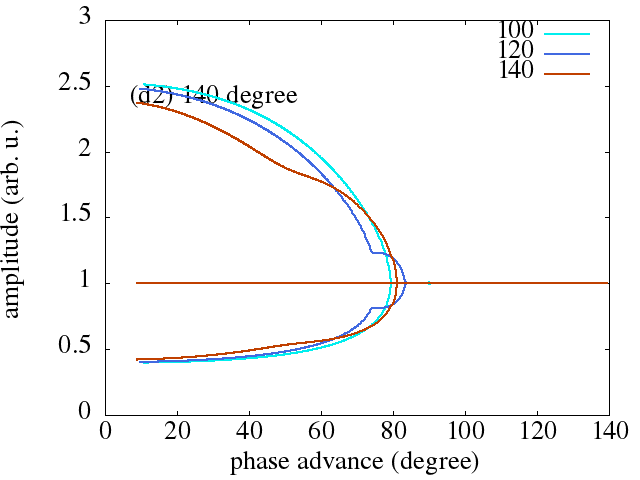}
	   \caption{The 3D envelope mode amplitudes as a function of 
	depressed transverse phase advance with $20$, $40$,
	$60$, $80$, $100$, $120$, and $140$ degree zero current 
	longitudinal phase advances
	for (a) $80$ degree, (b) $100$ degree, (c) $120$ degree,
	and (d) $140$ degree zero current transverse phase advances in
	a periodic quadrupole-RF channel.
	}
   \label{fd3damp}
\end{figure*}
\begin{figure*}[!htb]
   \centering
   \includegraphics*[angle=0,width=210pt]{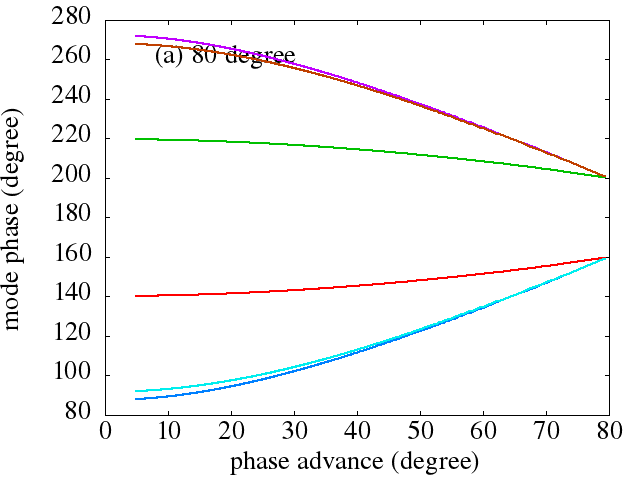} 
   \includegraphics*[angle=0,width=210pt]{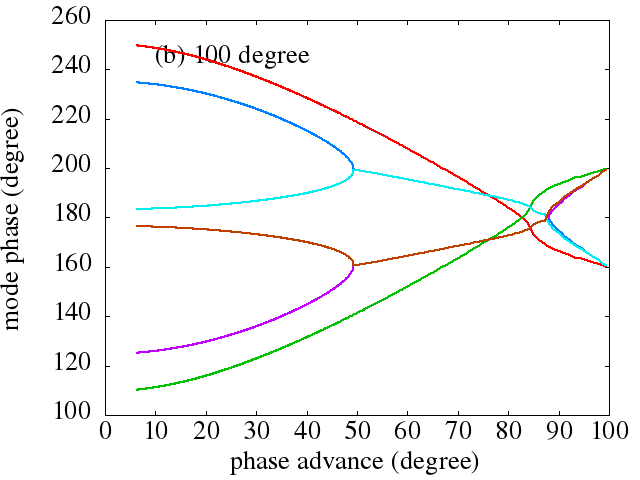} 
   \includegraphics*[angle=0,width=210pt]{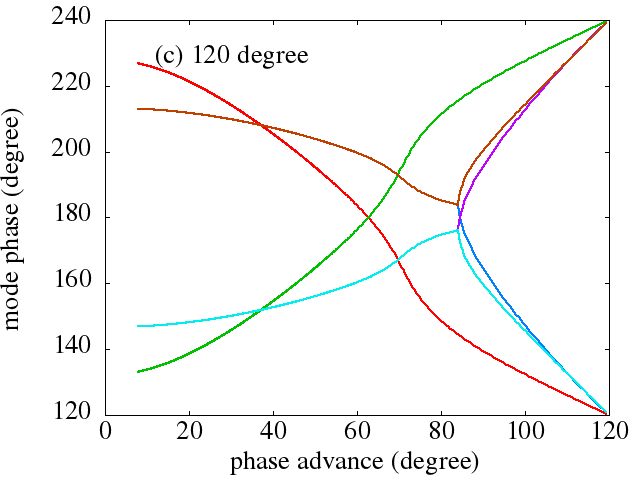} 
   \includegraphics*[angle=0,width=210pt]{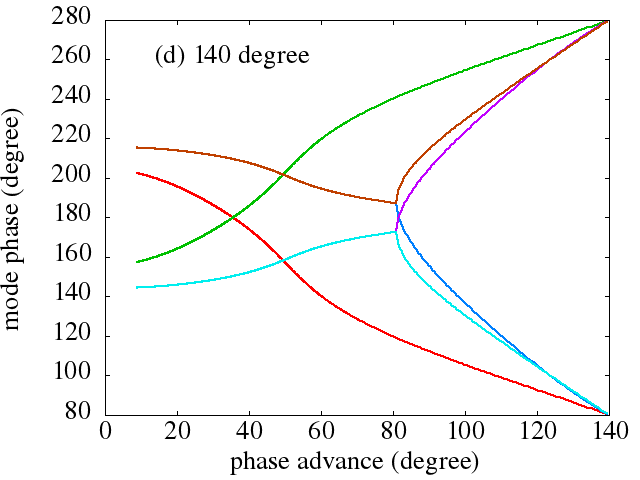} 
	  \caption{The 3D envelope mode phases as a function of 
	depressed transverse phase advance with (a) 
	$80$ degree, (b) $100$ degree, (c) $120$ degree,
	and (d) $140$ degree zero current longitudinal
	and transverse phase advances in a periodic quadrupole-RF channel.
	}
   \label{fd3dphase}
\end{figure*}
\begin{figure*}[!htb]
   \centering
   \includegraphics*[angle=0,width=210pt]{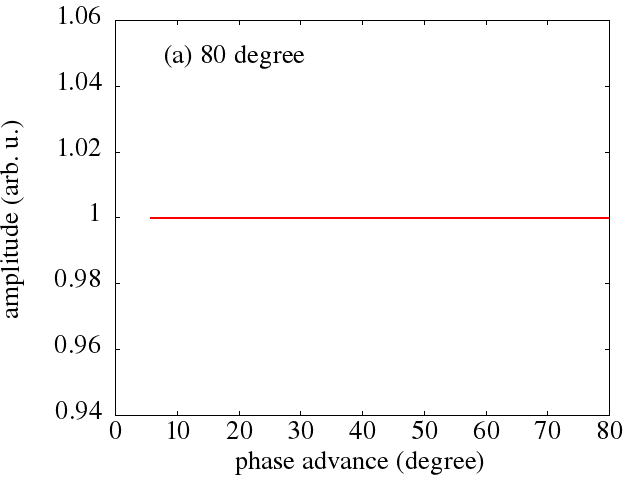}
   \includegraphics*[angle=0,width=210pt]{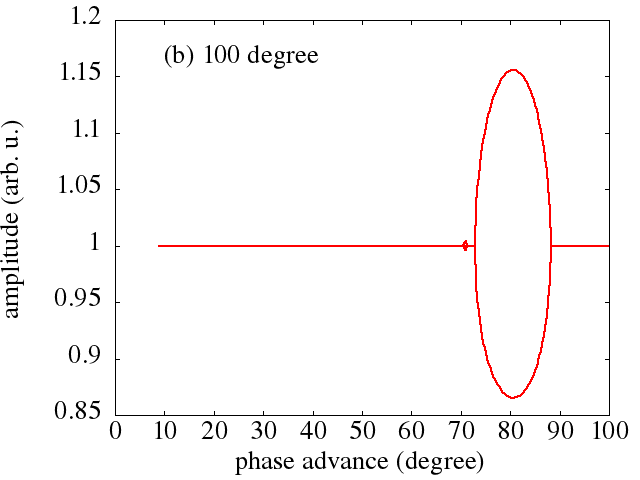}
   \includegraphics*[angle=0,width=210pt]{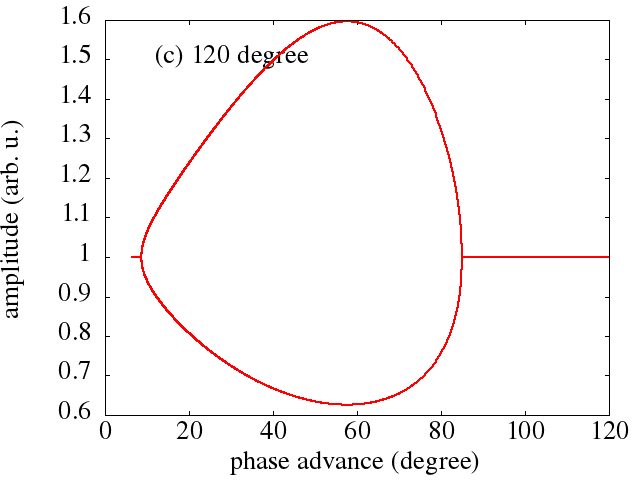}
   \includegraphics*[angle=0,width=210pt]{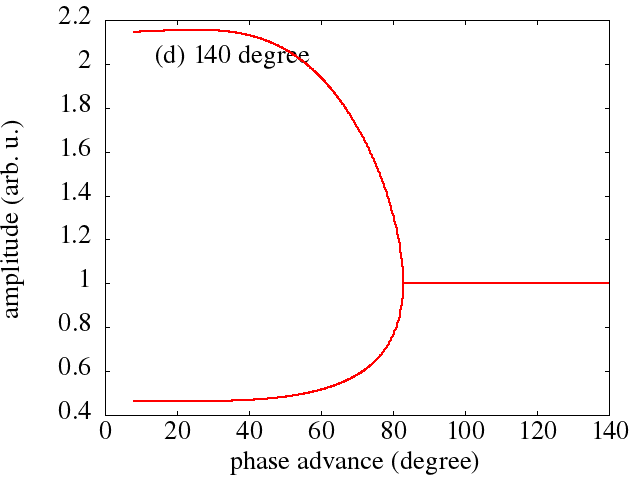}
	  \caption{The 2D envelope mode amplitudes as a function of 
	depressed transverse phase advance
	for (a) $80$ degree, (b) $100$ degree, (c) $120$ degree,
	and (d) $140$ degree zero current transverse phase advances in 
	a periodic quadrupole channel.  }
   \label{fd2damp}
\end{figure*}
\begin{figure*}[!htb]
   \centering
   \includegraphics*[angle=0,width=210pt]{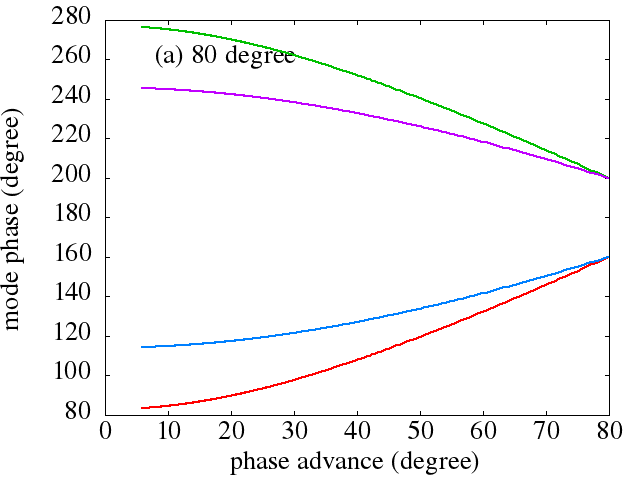}
   \includegraphics*[angle=0,width=210pt]{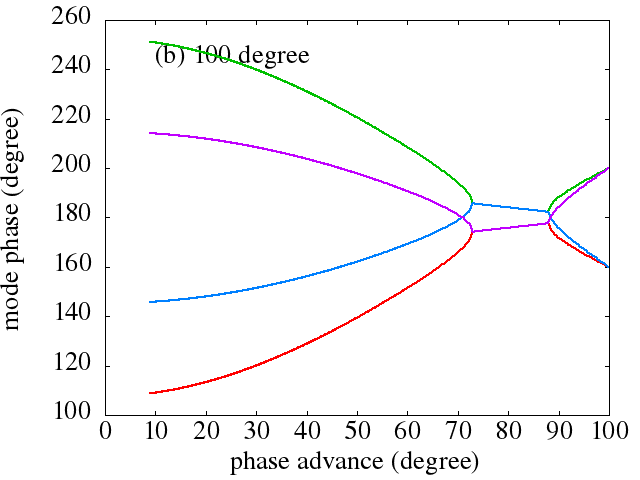}
   \includegraphics*[angle=0,width=210pt]{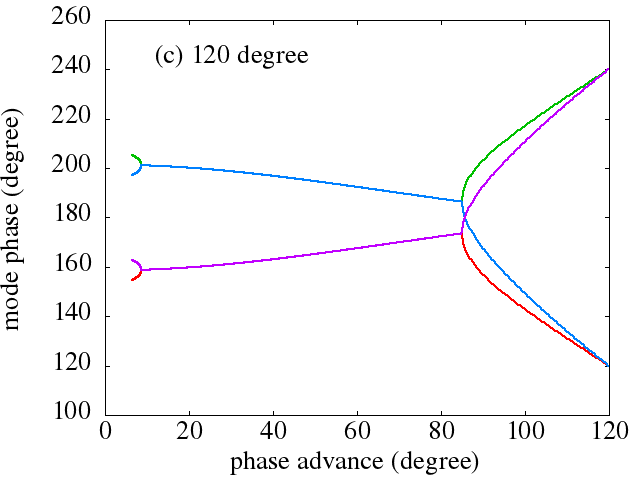}
   \includegraphics*[angle=0,width=210pt]{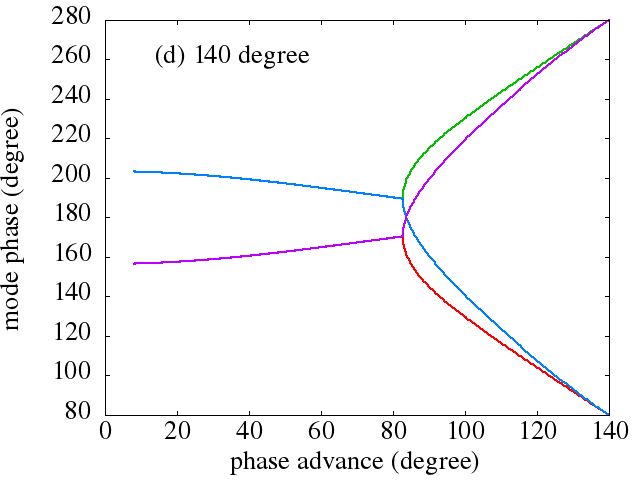}
	  \caption{The 2D envelope mode phases as a function of 
	depressed transverse phase advance
	for (a) $80$ degree, (b) $100$ degree, (c) $120$ degree,
	and (d) $140$ degree zero current transverse phase advances in 
	a periodic quadrupole channel.  }
   \label{fd2dphase}
\end{figure*}
\begin{figure}[!htb]
   \centering
   \includegraphics*[angle=0,width=210pt]{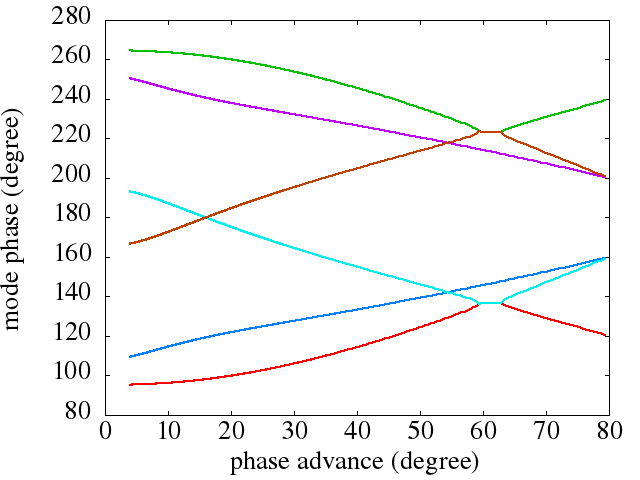} 
   \includegraphics*[angle=0,width=210pt]{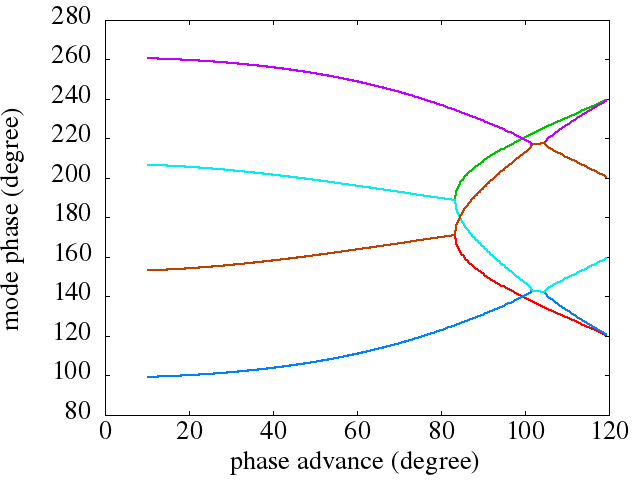} 
		  \caption{The 3D envelope mode phases as a function of 
	depressed transverse phase advance for zero current (a) 
	transverse $80$ degree and longitudinal $120$ degree, 
	(b) transverse $120$ degree and longitudinal $80$ degree
	phase advance in a periodic quadrupole channel. }
   \label{quad80120}
\end{figure}
A schematic plot of this periodic channel is shown in Fig.~\ref{fd3d}.
Each peroid of the channel consists of a $0.2$ meter focusing quadrupole,
	a $0.1$ meter RF focusing cavity, a $0.2$ meter defocusing
	quadrupole and another $0.1$
meter RF bunching cavity.
The total length of the period is $1.0$ meters.
Figures~\ref{fd3damp}-\ref{fd3dphase} show the 3D envelope mode amplitudes and phases
as a function of transverse depressed phase advance for different
zero current transverse and longitudinal phase advances.
As a comparison, we also show in Figs~\ref{fd2damp}-\ref{fd2dphase}
the 2D envelope mode amplitudes and phases as a function of the depressed 
phase advance for different zero current phase advances. Here, the 
2D periodic quadrupole channel has the same length of period as the 3D channel.
It is seen that in the 2D periodic quadrupole channel, the envelope instability 
occurs when the zero current phase advance is over $90$ degrees. 
There is no instability when the zero current phase advance is below $90$ 
degrees. In the
3D periodic quadrupole-RF channel, the envelope instability occurs even with
the zero current transverse phase advance $80$ degrees but the longitudinal 
phase advance beyond $100$ degrees in Fig.~\ref{fd3damp} (a2). 
There is no 
instability if both the transverse zero current phase advance and the longitudinal
zero current phase advance are below $90$ degrees.
	For the 3D envelope modes, when the longitudinal zero current
phase advance is below $90$ degrees and the transverse zero
current phase above $90$ degrees as shown in Fig.~\ref{fd3damp} (b1, c1, and d1), 
	the instability stopband width
increases with the increase of the zero current longitudinal phase advance.
For small longitudinal zero current phase advance (e.g. $20$
degrees), the 3D envelope modes instability stopband
is similar to that of the 2D envelope modes.
For the $100$ degree zero current transverse phase advance case, when
the zero current longitudinal phase advance is beyond $90$ degrees,
the stopband becomes more complicated and shows multiple stopbands.
For the transverse zero current $120$ and $140$ degree
phase advances, the instability stopbands do not change
significantly with the increase of zero current longitudinal phase
advance. 
This is probably due to the fact that when the transverse
zero current phase advance is beyond $100$ degrees, most parameter space 
(transverse depressed tune) below $90$ degrees becomes unstable
caused by the confluent resonance. Further increasing the
zero current longitudinal phase advance beyond $90$ degrees 
will not enlarge that stopband any more.


In the periodic transverse quadrupole focusing channel,
it is seen in Fig.~\ref{fd2dphase}, the 2D envelope instabilities are mainly due to
the confluent resonance between the two envelope modes when their phases 
become equal. This appears still to be valid in the 3D periodic
quadrupole-RF channel as shown in Fig.~\ref{fd3dphase}.

The 3D envelope instability shows asymmetry between the
transverse and the longitudinal direction in the 3D periodic quadrupole
and RF channel too. Figure~\ref{quad80120} shows the envelope mode phases as a 
function of depressed transverse phase advance for a case with zero current
$80$ degree transverse phase advance and $120$ degree longitudinal
phase advance, and a case with zero current $120$ degree transverse phase advance
and $80$ degree longitudinal phase advance. The envelope mode amplitudes 
are shown in Fig.~\ref{fd3damp} (a2 and c1) for this comparison. 
For the $80$ degree zero current transverse
phase advance, there is
only one major unstable region around $60$ degree depressed transverse 
phase advance
due to the confluent resonance.
For the $120$ degree zero current phase advance, there are two 
unstable regions due to two confluent resonances.

\begin{figure}[!htb]
   \centering
   \includegraphics*[angle=0,width=210pt]{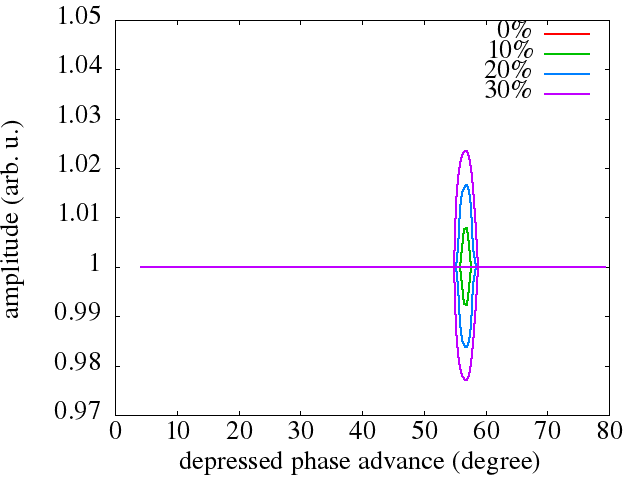} 
   \includegraphics*[angle=0,width=210pt]{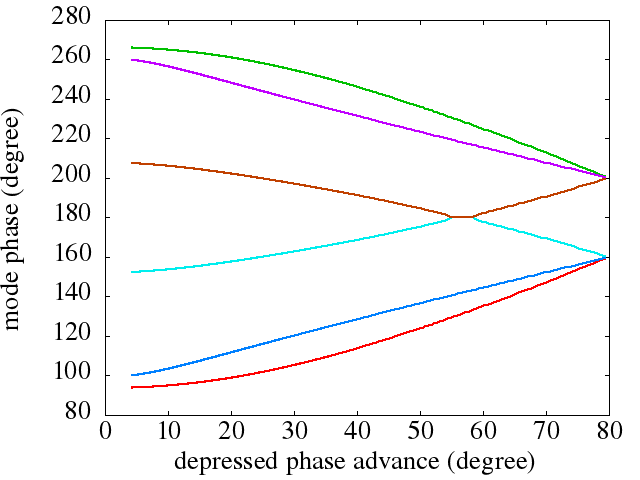}
	\caption{The 3D envelope mode (left) amplitudes and (right) phases as a 
function of the transverse depressed phase advance 
with $10\%$, $20\%$, and $30\%$ deviations from the original RF cavity setting
	in a periodic quadrupole-RF channel.}
   \label{quad80100}
\end{figure}

In the above periodic quadrupole and RF channel, we assumed that 
the two RF cavities have the same longitudinal focusing strength.
The longitudinal focusing period is half of the transverse
focusing period. This accounts for the absence of the envelope
instability for the zero current $80$ degree transverse phase 
advance and $100$ degree longitudinal phase in the periodic
quadrupole and RF channel.
The envelope instability stopband
is observed in the periodic solenoid and RF channel with the 
same zero current phase advances as shown in Fig.~\ref{sol3damp} (a2). 
The absence of instability for longitudinal zero current phase advance
$100$ degrees was also observed in 3D macroparticle simulations in
reference~\cite{ingo4}. Now, we break the symmetry of two RF longitudinal 
focusing cavities, the longitudinal focusing period becomes the
same as the transverse focusing period.
The envelope instability occurs for these zero
current phase advances in a periodic quadrupole and RF channel.
Figures~\ref{quad80100} show the envelope mode amplitudes and phases as a 
function of transverse depressed phase advances with about
$10\%$, $20\%$, and $30\%$ deviation from the original setting of the two
RF cavities (one cavity plus that percentage and the other one minus 
that percentage).
It is seen that as the asymmetry between the two RF cavity increases,
the instability stopband width also increases. 
Before breaking of the symmetry of two RF cavities, the longitudinal
phase advance per longitudinal period is $50$ degrees. After the breaking of the symmetry,
the longitudinal period becomes the same as the lattice period
and the phase advance becomes $100$ degrees. 
Such a zero current phase advance results in half integer parametric resonance
as shown in Fig.~\ref{quad80100}. 

\begin{figure}[!htb]
   \centering
   \includegraphics*[angle=0,width=210pt]{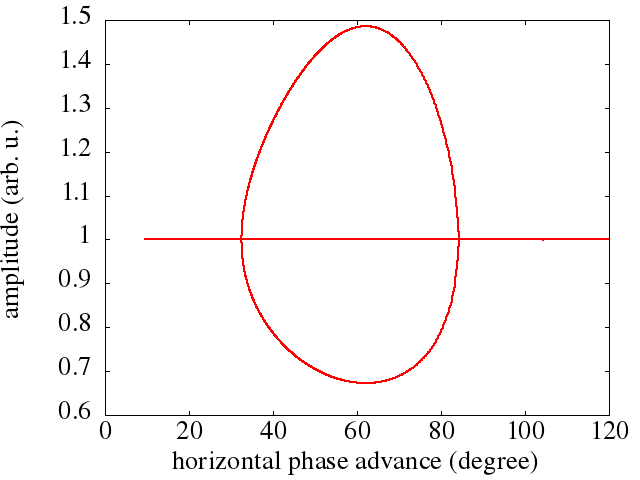} 
   \includegraphics*[angle=0,width=210pt]{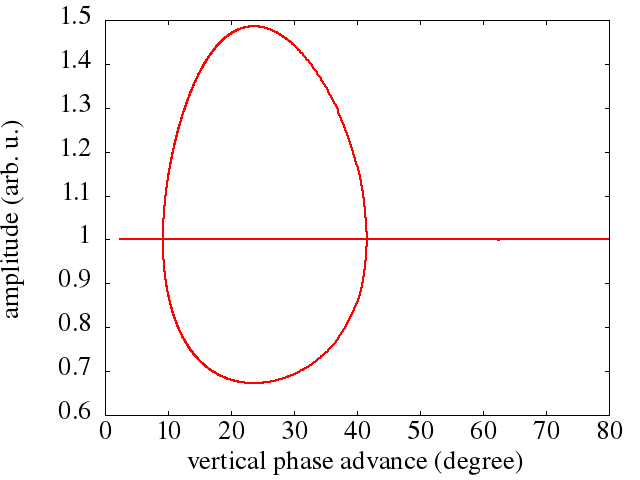} 
   \includegraphics*[angle=0,width=210pt]{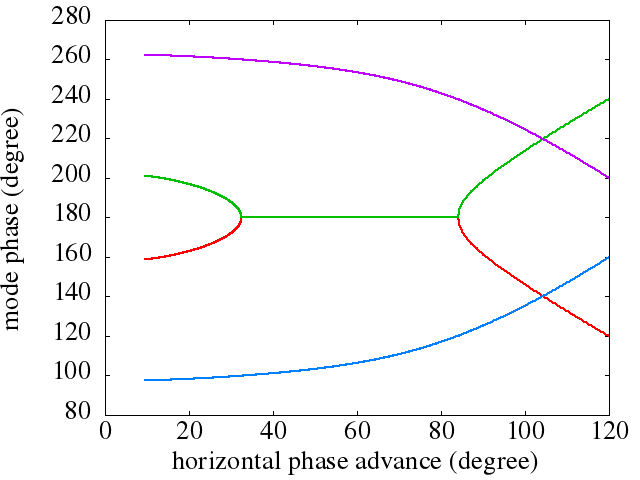} 
   \includegraphics*[angle=0,width=210pt]{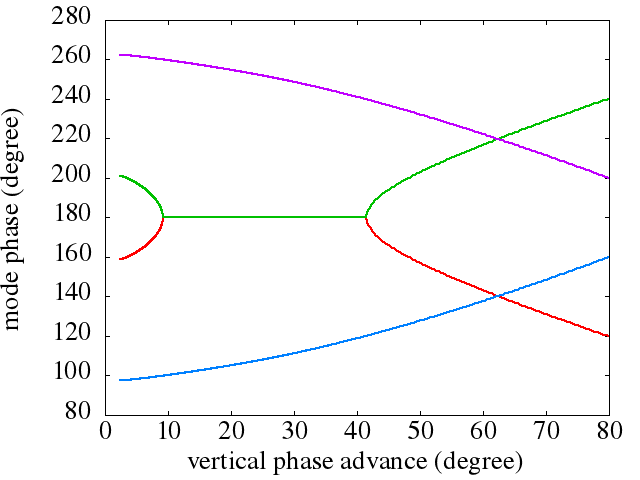} 
	\caption{The 2D envelope mode (top) amplitudes and (bottom) phases as a 
function of depressed phase advance with asymmetric zero current 
	phase advances ($80$ degrees in one direction and $120$ degrees
	in another direction) in a periodic quadrupole channel.}
   \label{fd2d80120}
\end{figure}

In above 3D periodic solenoid/quadrupole and RF transport channels, 
we have assumed that in transverse plane, the zero current phase advances in
horizontal direction and the vertical direction are the same.
Furthermore, the bunch has the same emittances in both horizontal and
vertical directions. This might imply a two-dimensional 
	transverse and longitudinal periodic system (i.e. $r-z$).
As a comparison, we also calculated the
envelope mode amplitudes and phases for a true
two-dimensional periodic quadrupole
channel with different zero
current phase advances in the horizontal and the vertical direction 
($120$ degrees
in the horizontal direction and $80$ degrees in the vertical direction). 
Figure~\ref{fd2d80120} shows the 2D envelope mode amplitudes and
phases as a function of the depressed horizontal and vertical phase advance.
Comparing the 2D envelope mode amplitudes and phases in above plot with those 
of the 3D envelope mode with the same zero current phase
	advances in Figs.~\ref{sol3damp} (a2) and \ref{fd3damp} (a2) 
($80$ degrees in transverse and $120$ in longitudinal)
	and Fig.~\ref{sol3damp} (c1) and \ref{fd3damp} (c1) 
($120$ degrees in transverse and $80$ in longitudinal),
we see that the 2D envelope instability shows somewhat similar structure to
the 3D envelope instability in a periodic solenoid-RF channel with
transverse zero current phase advance $80$ degrees and longitudinal phase advance $120$ degrees. The major instabilities in both cases are caused
by the half-integer parametric resonance.
The 3D envelope modes in a periodic quadrupole-RF channel shows quite 
different
instability stopband from the 2D envelope modes. Also the 3D
envelope instability in quadrupole channel is caused by the confluent
resonance while the 2D asymmetric envelope instability in the quadrupole
channel is
mainly caused by the half-integer parametric resonance.

\begin{figure}[!htb]
   \centering
   \includegraphics*[angle=0,width=210pt]{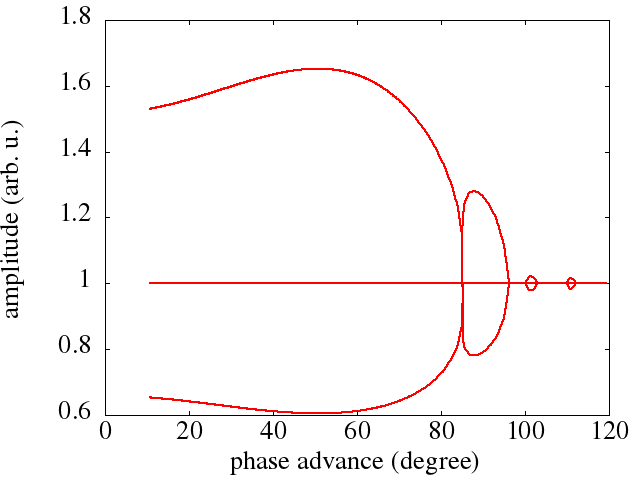} 
   \includegraphics*[angle=0,width=210pt]{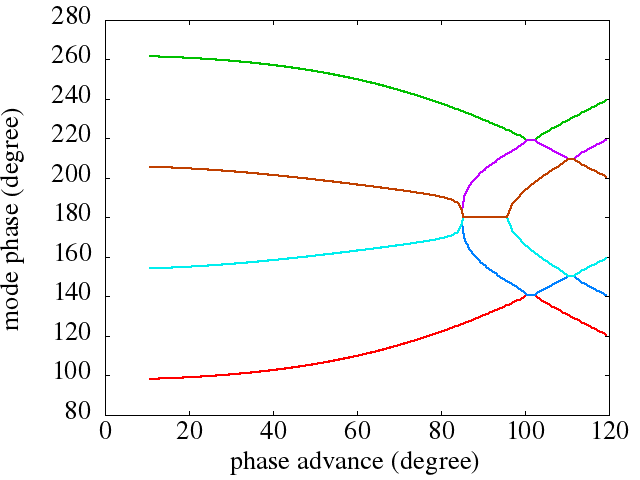} 
	\caption{The 3D envelope mode (left) amplitudes and (right) phases as a 
function of the horizontal depressed phase advances 
with zero current phase advances $120$ degrees in horizontal, $110$ in vertical,
and $80$ in longitudinal direction
	in a periodic quadrupole-RF channel.}
   \label{quad3d11012080}
\end{figure}

We also explored 3D envelope instabilities with non-equal
transverse zero current phase advances in the horizontal
direction and the vertical direction.
Figure~\ref{quad3d11012080} shows the 3D envelope mode amplitudes and phases as a 
function of the depressed horizontal tune with zero current phase advance 
$120$ degrees
in the horizontal direction, $110$ degrees in the vertical direction,
and $80$ degrees in the longitudinal direction in the periodic quadrupole
and RF channel.
Comparing the above figure with the zero current $120$ degree transverse phase 
	advance and $80$ degree longitudinal phase advance case in Fig.~\ref{fd3damp} (c1),
we see that 3D instability stopband from the nonequal transverse focusing 
becomes broader. 
Instead of one major instability stopband and a minor stopband in
the equal transverse phase advance case, now there are four
stopbands (two major stopbands and two minor stopbands) for the transverse
$120$ and $110$ degree phase advances.
Besides the confluent resonance,
there also appears a half-integer parametric resonance when the transverse
symmetry is broken. Breaking the transverse symmetry results in
more resonances of these envelope modes.
	This suggests that keeping the same zero
current phase advance in both the horizontal and the vertical directions
might help reduce the parameter region of the envelope instability.

\section{Conclusions}
In this paper, we proposed a three-dimensional envelope instability
model to study the instability for a bunched beam
in a periodic solenoid and RF focusing channel and a periodic quadrupole and RF focusing channel.
This study showed that when the transverse zero current phase advance 
is below $90$ degrees, the beam envelope can still become unstable if
the longitudinal zero current phase advance is beyond $90$ degrees.
For the transverse zero current phase advance beyond $90$ degrees, 
the instability stopband becomes broader with the increase of 
longitudinal focusing strength and even shows different
structure from the 2D case 
when the longitudinal zero current phase advance is
beyond $90$ degrees. 

The 3D envelope instability shows asymmetry between the longitudinal
focusing and the transverse focusing.
The instability shows broader stopband when the transverse zero current
phase advance is beyond $90$ degrees than that when the longitudinal
zero current phase advance is beyond $90$ degrees.
In the 3D periodic quadrupole and RF channel, for the transverse 
zero current phase advance $80$ degree, the envelope modes stay stable 
for the longitudinal $100$ degree zero current phase advance due to
the symmetry of two longitudinal focusing RF cavities. Breaking the symmetry
of two cavities results in the envelope instability with a finite stopband.
Breaking the horizontal and vertical focusing symmetry in the transverse 
plane also increases the 
envelope instability stopband width. This suggests that a
more symmetric accelerator lattice design might help reduce the parameter space of the
envelope instability.

\section*{ACKNOWLEDGEMENTS}
Work supported by the U.S. Department of Energy under Contract No. DE-AC02-05CH11231.
We would like to thank Dr. R. D. Ryne for the use of his 2D and 3D envelope codes.  This research used computer resources at the National Energy Research
Scientific Computing Center.


\begin{thebibliography}{9}   


\bibitem{ingo1}I. Hofmann, L. J. Laslett, L. Smith, and I. Haber, Part. Accel. 13, 145 (1983).
\bibitem{jurgen}J. Struckmeier and M. Reiser, Part. Accel. 14, 227 (1984).
\bibitem{ingo3}I. Hofmann, Phys. Rev. E 57, 4713 (1998). 
\bibitem{davidson}R. C. Davidson, H. Qin, and G. Shvets, Phys. Plasmas 7, 1020 (2000). 
\bibitem{okamoto}H. Okamoto and K. Yokoya, Nucl. Instrum. Methods Phys. Res., A 482, 51 (2002). 
\bibitem{fedotov0}A. V. Fedotov and I. Hofmann, Phys. Rev. Spec. Top.-Accel. Beams 5, 024202 (2002). 
\bibitem{fedotov}A. V. Fedotov, I. Hofmann, R. L. Gluckstern, and H. Okamoto, Phys. Rev. Spec. Top.-Accel. Beams 6, 094201 (2003). 
\bibitem{lund}S. M. Lund and B. Bukh, Phys. Rev. Spec. Top.-Accel. Beams 7, 024801 (2004). 
\bibitem{tiefenback}M. Tiefenback, ``Space charge limits on the transport of ion beams in a long alternating gradient system,'' Ph.D. thesis, Lawrence Berkeley National Laboratory Report LBL-22465, 1986.
\bibitem{gilson}E. P. Gilson, M. Chung, R. C. Davidson  P. C. Efthimion, and R. Majeski, Phys. Rev. Spec. Top.-Accel. Beams 10, 124201 (2007). 
\bibitem{groening}L. Groening, W. Barth, W. Bayer, G. Clemente, L. Dahl, P. Forck, P. Gerhard, I. Hofmann, M. S. Kaiser, M. Maier, S. Mickat, and T. Milosic, Phys. Rev. Lett. 102, 234801 (2009). 
\bibitem{jeon1}D. Jeon, L. Groening, and G. Franchetti, Phys. Rev. Spec. Top.-Accel. Beams 12, 054204 (2009). 
\bibitem{fukushima}K. Fukushima, K. Itoa, H. Okamoto, S. Yamaguchi, K. Moriya, H. Higaki, T. Okano, and S. M. Lund, Nucl. Instrum. Methods Phys. Res., A 733, 18 (2014). 
\bibitem{li0}C. Li and Y. L. Zhao, Phys. Rev. ST Accel. Beams 17, 124202 (2014).
\bibitem{li}C. Li and Q. Qin, Phys. Plasmas 22, 023108 (2015).
\bibitem{ingo2}I. Hofmann and O. Boine-Frankenheim, Phys. Rev. Lett. 115, 204802 (2015). 
\bibitem{jeon2}D. Jeon, J. H. Jang, and H. Jin,
	Nucl. Instrum. Methods Phys. Res., A 832, 43 (2016). 
\bibitem{oliver}O. Boine-Frankenheim, I. Hofmann, and J. Struckmeier,
Phys. Plasmas 23, 090705 (2016).
\bibitem{ingoprab17}I. Hofmann and O. Boine-Frankenheim, 
Phys. Rev. Accel. Beams 20, 014202 (2017).
\bibitem{ito}K. Ito, H. Okamoto, Y. Tokashiki, and K. Fukushima, 
Phys. Rev. Accel. Beams 20, 064201 (2017).
\bibitem{yuan}Y. Yuan, O. Boine-Frankenheim, and I. Hofmann,
Phys. Rev. Accel. Beams 20, 104201 (2017).
\bibitem{ingo4}I. Hofmann and O. Boine-Frankenheim, Phys. Rev. Lett. 118, 
	114803, (2017).
\bibitem{ingobook}I. Hofmann, ``Space Charge Physics for Particle Accelerators,'' Springer, 2017.
\bibitem{kv}I. Kapchinskiy and V. Vladimirskiy, 2nd Conf. High Energy Accel., p. 274, CERN, Geneva, 1959.
\bibitem{bongardt}K. Bongardt, M. Pabst and A. Letchford, in Proc. 1998 Linac Conf., p. 824, 1999.
\bibitem{allen}C. Allen, in Proc. of the 2nd ICFA Advanced Accelerator 
Workshop, Ed. J. Rosenzweig and L. Serafini, p. 173, World Scientific, 2000.
\bibitem{qiang0}J. Qiang and R. D. Ryne, Phys. Rev. ST Accel. Beams 3,
064201 (2000).
\bibitem{comunian}M. Comunian, A. Pisent, A. Bazzani, G. Turchetti, and S. Rambaldi, Phys. Rev. ST Accel. Beams 4,
124201 (2001).
\bibitem{sacherer}F. J. Sacherer, IEEE Trans. Nucl. Sci. 18, 1101 (1971).
\bibitem{ryne}R. Ryne, Los Alamos Report No. LA-UR-95-391;; http://xxx.lanl.gov/abs/acc-phys/9502001.
\end{thebibliography}
\end{document}